\documentclass[12pt,a4paper]{article} 
\usepackage{jheppub}
\usepackage{enumerate}
 \usepackage{epstopdf}
\usepackage{float} 
\usepackage[caption = false]{subfig}
\usepackage{amsmath} 
\usepackage{wasysym} 
\usepackage{mathrsfs} 
\usepackage{graphicx} 
\usepackage{amsfonts} 
\usepackage{amsbsy} 
\usepackage{amscd} 
\usepackage{multirow} 
\usepackage{tikz}
\usepackage{color}
\usepackage{slashed}
\usepackage{array}
\usepackage{tablefootnote}
\usetikzlibrary{arrows,positioning,shapes.geometric} 
\usepackage{color}
 
\definecolor{myred}{rgb}{0.6,0,0}
\definecolor{myblue}{rgb}{0,0.2,0.4}
\definecolor{mygreen}{rgb}{0,0.9,0.1}
\definecolor{hc}{rgb}{.9,0.1,0.7}
\definecolor{hcout}{rgb}{.9,0.7,0.9}
\definecolor{Orange}{rgb}{1.,0.65,0.}

\newcommand{\re}[1]{{\color{red!80!black}{#1}}}

\numberwithin{equation}{section}
\numberwithin{figure}{section}
\numberwithin{table}{section}

\newcommand{\be}{\begin{equation}}
\newcommand{\ee}{\end{equation}}
\newcommand{\bea}{\begin{eqnarray}}
\newcommand{\eea}{\end{eqnarray}}

\newcommand{\nl}{\nonumber \\}

\newcommand{\newc}{\newcommand}
\newc{\bi}{\begin{itemize}}
\newc{\ei}{\end{itemize}}

\newc{\ra}{\rightarrow}
\newc{\sq}   {\mbox{$\wt{q}$}}
\newc{\msq}  {\mbox{$m_{\sq}$}}
\newc{\gl}   {\mbox{$\wt{g}$}}
\newc{\mgl}  {\mbox{$m_{\gl}$}}

\newc{\wt}{\widetilde}

\newc{\ifb}{\mbox{${\rm fb}^{-1}$}}

\newc{\del}{\delta}

\title{Scalar dark matter search from the extended $\nu$THDM } 
\author[a,b]{Seungwon Baek,}
\emailAdd{sbaek@korea.ac.kr}
\affiliation[a]{School of Physics, KIAS, Seoul 02455, Korea}
\affiliation[b]{Department of Physics, Korea University, Seoul 02841, Korea}
\author[a]{Arindam Das}
\emailAdd{arindam@kias.re.kr}
\author[a]{and Takaaki Nomura}
\emailAdd{nomura@kias.re.kr}

\abstract{ We consider a neutrino Two Higgs Doublet Model ($\nu$THDM) in which neutrinos obtain {\it naturally} small Dirac
masses from the soft symmetry breaking of a global $U(1)_X$ symmetry. We extended the model so the soft term
is generated by the spontaneous breaking of $U(1)_X$ by a new scalar field. The symmetry breaking pattern can also
stabilize a scalar dark matter candidate. After constructing the model,
we study the phenomenology of the dark matter: relic density, direct and indirect detection.
}

\preprint{\re{\today}}


\begin{document}

\maketitle
%%%%%%%%%%%%%%%%%%%%%%%%%%%%%%%%%%%%%%%%%%%%%%%%%%
\section{Introduction}
\label{sec:into}
%%%%%%%%%%%%%%%%%%%%%%%%%%%%%%%%%%%%%%%%%%%%%%%%%%%%%%%%%%%%%%%%%%%%%%%%%%%%%
The existence of the tiny neutrino mass can be naturally explained by the seesaw mechanism \cite{Minkowski:1977sc, Yanagida:1980xy, Schechter:1980gr, Sawada:1979dis, GellMann:1980vs,Glashow:1979nm, Mohapatra:1979ia} which extends the Standard Model (SM) through Majorana type Right Handed Neutrinos (RHNs). 
As a result the SM light neutrinos become Majorana particles. Alternatively there is a simple model, 
neutrino Two Higgs Doublet Model ($\nu$THDM) \cite{Davidson:2009ha, Wang:2006jy}, which can generate the Dirac mass term for the
 light neutrinos as well as for the other fermions in the SM. In this model we have two Higgs doublets; one is the same as the SM-like Higgs doublet and the other one is having a small VEV $(\mathcal{O}(1))$ eV
 to explain the tiny neutrino mass correctly. Due to this fact, the neutrino Dirac Yukawa coupling could be order 1. It has been discussed in \cite{Davidson:2009ha} that a global softly broken $U(1)_X$ symmetry
can forbid the Majorana mass terms of the RHNs;  a hidden $U(1)$ gauge symmetry can be also applied to realize $\nu$THDM as in ref.~\cite{Nomura:2017wxf}. In this model all the SM fermions obtain Dirac mass terms via Yukawa interactions with the SM-like Higgs doublet $(\Phi_2)$ whereas only the neutrinos get 
Dirac masses through the Yukawa coupling with the other Higgs doublet $(\Phi_1)$. Another scenario of the generation of Dirac neutrino mass through a dimension five operator has been studied in \cite{CentellesChulia:2018gwr}. The corresponding Yukawa interactions of the Lagrangian can be written as 
\bea
\mathcal{L}_{Y}=-\overline{Q}_L Y^u \widetilde{\Phi}_2 u_R -\overline{Q}_L Y^d \Phi_2 d_R 
-\overline{L}_L Y^e \Phi_2 e_R -\overline{L}_L Y^\nu \widetilde{\Phi}_1 \nu_R +\rm{H. c.}
\label{Yuk1}
\eea
where $\widetilde{\Phi}_i = i \sigma_2 \Phi_i^* (i=1,2)$, $Q_L$ is the SM quark doublet, $L_L$ is the SM lepton doublet, $e_R$ is the right handed charged lepton, $u_R$ is the right handed up-quark, $d_R$ is the right handed down-quark and $\nu_R$ are the RHNs.
The $\Phi_1$ and $\nu_R$ are assigned with the global charge  $3$ under the $U(1)_X$ group. The global symmetry forbids the
Majorana mass term between the RHNs. In the original model~\cite{Davidson:2009ha}, the global symmetry is softly broken by the mixed mass 
term between $\Phi_1$ and $\Phi_2$ $(m_{12}^2 \Phi_1^\dagger \Phi_2)$ such that a small VEV is obtained by seesaw-like formulas
\bea
v_1 =\frac{m_{12}^2 v_2}{M_A^2},
\eea
where $M_A$ is the pseudo-scalar mass in \cite{Davidson:2009ha}. If $M_A \sim 100$ GeV and $m_{12} \sim {\cal O}(100)$
keV then $v_1$ can be obtained as  $\mathcal{O}(1)$ eV. In the paper~\cite{Baek:2016wml}, the model is extended to
include singlet scalar $S$ which breaks the $U(1)_X$ symmetry. The soft term $m_{12}^2$ is identified with 
$\mu \langle S \rangle$ where $\mu$ is the Higgs mixing term, $\mu \Phi_1^\dagger \Phi_2 S + h.c.$. 
It has been studied in~\cite{Baek:2016wml} 
that an SM singlet fermion being charged under $U(1)_X$ could be a potential DM candidate.

 In this paper we extend the model with a natural scalar Dark Matter (DM)
candidate $(X)$. In this model the global $U(1)_X$ symmetry is spontaneously broken down to $Z_2$ symmetry by VEV of a new singlet scalar $S$. 
The remnant of the $Z_2$ symmetry makes the DM candidate stable.
The $Z_2$ symmetry would be broken by quantum gravity effect and DM would decay via effective interaction \cite{Mambrini:2015sia}. This can be avoided if the $U(1)_X$ is a remnant of local symmetry at a high energy scale 
and we assume the $Z_2$ symmetry is not broken.
 A CP odd component  of $S$
becomes the Goldstone boson and hence we study the DM annihilation from this model and compare with the current experimental sensitivity. 

The papers is organized as follows. In Sec.~\ref{Model} we describe the model. In Sec.~\ref{DMP} we discuss the DM phenomenology and finally in Sec.~\ref{Conc} we conclude.

%%%%%%%%%%%%%%%%%%%%%%%%%%
\section{The Model}
\label{Model}
%%%%%%%%%%%%%%%%%%%%%%%%%%
We discuss the extended version of the model in \cite{Davidson:2009ha} with a scalar field $(X)$. We write the scalar and the RHN sectors of the particle content in Tab.~\ref{tab1} 
\begin{table}[h] 
\centering
\begin{tabular}{|c||c|c|c|c||c|}\hline\hline  
&\multicolumn{4}{c||}{Scalar Fields} & \multicolumn{1}{c|}{New Fermion} \\\hline
& ~$\Phi_1$~ & ~$\Phi_2$~ & ~$S$ ~ & ~$X$~ & ~$\nu_R$    \\\hline 
$SU(2)_L$ & $\bf{2}$  & $\bf{2}$ & $\bf{1}$ & $\bf{1}$ & $\bf{1}$   \\\hline 
$U(1)_Y$ & $\frac12$ & $\frac12$  & $0$ & $0$ & $0$    \\\hline
 $U(1)_X$ & $3$ & $0$   & $3$ & $1$  & $3$     \\\hline
%%%
\end{tabular}
\caption{Scalar fields and new fermion in our model.} %where $\nu_R$ is Majorana type.}
\label{tab1}
\end{table}
The gauge singlet Yukawa interaction between the lepton doublet $(L_L)$, the doublet scalars $(\Phi_1, \Phi_2)$ and the RHNs $(\nu_R)$ can be written as 
\bea
\mathcal{ L}  &\supset & - Y_{ij}^e \bar L_{L_{i}} \Phi_2 e_{Rj} - Y^\nu_{ij} \bar L_{L_{i}}  \tilde \Phi_1 \nu_{Rj} + \rm{H.c}.
\label{eq:Yukawa}
\eea
We assume that the Yukawa coupling constants $Y_{ij}^e$ and $Y_{ij}^\nu$ are real. The scalar potential can be written by
\bea
V(\Phi_1, \Phi_2, S) &= & - m_{11}^2 \Phi^\dagger_1 \Phi_1 - m_{22}^2 \Phi_2^\dagger \Phi_2 - m_{S}^2 S^\dagger S + M_X^2 X^\dagger X - (\mu \Phi^\dagger_1 \Phi_2 S + h.c.) \nonumber \\
&& + \lambda_1 (\Phi_1^\dagger \Phi_1)^2 + \lambda_2 (\Phi_2^\dagger \Phi_2)^2 + \lambda_3 (\Phi_1^\dagger \Phi_1)( \Phi_2^\dagger \Phi_2) 
+ \lambda_4 (\Phi_1^\dagger \Phi_2)( \Phi_2^\dagger \Phi_1)  \nonumber \\
&&+ \lambda_S (S^\dagger S)^2+ \lambda_{1S} \Phi_1^\dagger \Phi_1 S^\dagger S + \lambda_{2S} \Phi_2^\dagger \Phi_2
S^\dagger S +\lambda_X (X^\dagger X)^2 + \lambda_{1X} \Phi_1^\dagger \Phi_1 X^\dagger X  \nonumber  \\
&& + \lambda_{2X} \Phi_2^\dagger \Phi_2 X^\dagger X+ \lambda_{SX} S^\dagger S X^\dagger X  %(\lambda_{12X} \Phi_1^\dagger \Phi_2 X X + \rm H.c.)\nonumber \\
 - (\lambda_{3X} S^\dagger X X X + \rm{H.c.}) ,
\label{eq:potential}
\eea
The Dirac mass terms of the neutrinos are generated by the small VEV of $\Phi_1$. According to \cite{Davidson:2009ha, Wang:2006jy} we assume that the VEV of $\Phi_1$ is much smaller than the electroweak scale. 
The vacuum stability analysis of a general scalar potential has been studied in \cite{Kannike:2016fmd}.
Additionally, a remaining $Z_3$ symmetry is also involved when $U(1)_X$  is broken by non-zero VEV of S. 
 Here $X$ is the only $Z_3$ charged stable (scalar) particle and as a result $X$ could be considered as a potential Dark Matter
  (DM) candidate. The mass term $M_X$ of $X$ in Eq.~\ref{eq:potential} is positive definite which forbids $X$ to get VEV and as a result the
  $Z_3$ symmetry promotes the stability of $X$ as a DM candidate. It has already been discussed in \cite{Baek:2016wml} that  a
  CP-odd component in $S$ becomes massless Goldstone boson. Then we write scalar fields as follows
\bea
\Phi'_1 &=& \begin{pmatrix} \phi^+_1 \\ \frac{1}{\sqrt{2}} (v_1 + h_1 + i a_1) \end{pmatrix}, \quad
 \Phi_2 = \begin{pmatrix} \phi^+_2 \\ \frac{1}{\sqrt{2}} (v_2 + h_2 + i a_2) \end{pmatrix}, \\ %\quad X' = \frac{1}{\sqrt{2}} (X_R + i X_I),\nonumber \\
   \quad X &=& X' e^{ i\frac{a_S}{2v_S}},  \quad \Phi_1 = \Phi'_1 e^{ i\frac{a_S}{v_S}}, ~~S = \frac{1}{\sqrt{2}} r_S e^{ i\frac{a_S}{v_S}},
  \eea
where $r_S = \rho + v_S$. We assume $X$ does not develop a VEV while the VEVs of $\Phi_1$, $\Phi_2$ and $S$ are obtained by  requiring the stationary conditions $\partial V(v_1,v_2,v_S)/\partial v_i =0$ following
\bea
 -2 m_{11}^2 v_1 + 2 \lambda_1 v_1^3 + v_1 (\lambda_{1S} v_S^2 + \lambda_3 v_2^2 + \lambda_4 v_2^2) - \sqrt{2} \mu v_2  v_S &=&0, \nonumber \\
-2 m_{22}^2 v_2 + 2 \lambda_2 v_2^3 + v_2 (\lambda_{2S} v_S^2 + \lambda_3 v_1^2 + \lambda_4 v_1^2) - \sqrt{2} \mu v_1 v_S &=& 0, \nonumber \\
 -2 m_{S}^2 v_S + 2 \lambda_S v_S^3 + v_S (\lambda_{1S} v_1^2 + \lambda_{2S} v_2^2 ) - \sqrt{2} \mu v_1 v_2 &=& 0. 
\label{eqstn}
\eea
We then find that these conditions can be satisfied with $v_1 \simeq \mu \ll \{ v_2, v_S \}$ and SM Higgs VEV is given
as $v \simeq v_2 \simeq 246$ GeV. From the first one of the Eq.~\ref{eqstn} we find that $v_1$ is proportional to and of the same order with $\mu$ such that
\bea
v_1 \simeq \frac{\sqrt{2} \mu v_2 v_S}{\lambda_{1S} v_S^2 +(\lambda_3+\lambda_4) v_2^2 -2 m_{11}^2}.
\eea
The small order of $v_1 (\sim \mu)$ is required to keep $v_2$ and $v_S$ in the electroweak scale. Considering the
neutrino mass scale as $m_\nu \sim 0.1$ eV, the value of $\mu/v_2$ should be small such as $\mu/v_2 \sim {\mathcal
  O}(10^{-12})$ ensuring $Y^\nu$ as ${\mathcal O}(1)$ such that $m_e/v_2 \sim {\mathcal O}(10^{-6})$.  Hence $v_1$ is
considered to be smaller than the other VEVs. It also interesting to notice that $\mu=0$ restores the symmetry of the
Lagrangian hence a technically natural small value of $\mu$ is
acceptable \cite{tHooft:1979rat,Baek:2014sda}.
 It is also interesting to notice that $\mu=0$ enhances the  symmetry of the Lagrangian in the sense that we can assign 
 arbitrary $U(1)_X$ charge to $\Phi_1$, which ensures the radiative  generation of the $\mu$-term is proportional to $\mu$ itself. Hence a small
 value of $\mu$ is technically natural \cite{tHooft:1979rat,Baek:2014sda}.
Now we identify mass spectra in the scalar sector.

{\tt Charged scalar:} In this case we calculate the mass matrix in the basis $(\phi_{1}^{\pm}, \phi_{2}^{\pm})$ where $\phi_1^\pm $ is approximately physical charged scalar while $\phi_2^\pm$ is approximately NG boson absorbed by $W^\pm$ boson.
In the following we write physical charged scalar field as $H^\pm \simeq \phi^\pm_1$.
The charged scalar mass matrix can be written as 
\begin{equation}
M^2_{H^\pm} = \begin{pmatrix} \frac{v_2 (\sqrt{2} \mu v_S - \lambda_4 v_1 v_2 )}{2v_1} & - \frac{1}{2} (\sqrt{2}\mu v_S - \lambda_4 v_1 v_2)  \\ 
- \frac{1}{2} (\sqrt{2}\mu v_S - \lambda_4 v_1 v_2) & \frac{v_1 (\sqrt{2} \mu v_S - \lambda_4 v_1 v_2)}{2 v_2}  \end{pmatrix} 
 \simeq  \begin{pmatrix} \frac{v_2 (\sqrt{2} \mu v_S - \lambda_4 v_1 v_2 )}{2v_1}& 0 &  \\ 0 & 0 &  \end{pmatrix}.
\end{equation}
The charged Higgs mass can be written as 
\bea
m_{H^{\pm}}^2 \simeq \frac{v_2 (\sqrt{2} \mu v_S-\lambda_4 v_1 v_2)}{2 v_1}.
\eea

{\tt CP-even neutral scalar:} 
In the case of CP-even scalar all three components are physical. Hence the mass matrix can be written in the basis of $(h_1,h_2, \rho)$ as
\begin{align}
M^2_H &= \begin{pmatrix} 2 \lambda_1 v_1^2 + \frac{\mu v_2 v_S}{\sqrt{2} v_1} & (\lambda_3 + \lambda_4) v_1 v_2 - \frac{\mu v_S}{\sqrt{2}} & \lambda_{1S} v_1 v_S - \frac{\mu v_2}{\sqrt{2}} \\ 
(\lambda_3 + \lambda_4) v_1 v_2 - \frac{\mu v_S}{\sqrt{2}} & 2 \lambda_2 v_2^2 + \frac{\mu v_1 v_S}{\sqrt{2} v_2} & \lambda_{2S} v_2 v_S - \frac{\mu v_1}{\sqrt{2}} \\ 
\lambda_{1S} v_1 v_S - \frac{\mu v_2}{\sqrt{2}}  & \lambda_{2S} v_2 v_S - \frac{\mu v_1}{\sqrt{2}} & 2 \lambda_S v_S^2 + \frac{\mu v_1 v_2}{\sqrt{2} v_S} \end{pmatrix}  \nonumber \\
& \simeq \begin{pmatrix} \frac{ \mu v_2 v_S }{\sqrt{2} v_1} & 0 & 0 \\ 0 & 2 \lambda_2 v_2^2 & \lambda_{2S} v_2 v_S \\ 0 & \lambda_{2S} v_2 v_S & 2 \lambda_{S} v_S^2 \end{pmatrix}.
\end{align}
We find that all the masses of the mass eigenstates, $H_i (i=1,2,3)$, are at the electroweak scale and the mixings
between $h_1$ and other components are negligibly small while
the $h_2$ and $\rho$ can have sizable mixing. The mass eigenvalues and the mixing angle for $h_2$ and $\rho$ system can be given by
\begin{align}
& m_{H_2,H_3}^2 = \frac{1}{2} \left[ m_{22}^2 + m_{33}^2 \mp \sqrt{(m_{22}^2-m_{33}^2)^2 + 4 m_{23}^4} \right], \\
& \tan 2 \theta = \frac{-2 m_{23}^2}{m_{22}^2 - m_{33}^2}, \\
& m_{22}^2 = 2 \lambda_2 v_2^2, \quad m_{33}^2 = 2 \lambda_{S} v_S^2, \quad m_{23}^2 = \lambda_{2S} v_2 v_S.
\end{align}
Hence the mass eigenstates are obtained as 
\begin{equation}
\begin{pmatrix} H_1 \\ H_2 \\ H_3 \end{pmatrix} \simeq \begin{pmatrix} 1 & 0 & 0 \\ 0 & \cos \theta & - \sin \theta \\ 0 & \sin \theta & \cos \theta \end{pmatrix} \begin{pmatrix} h_1 \\ h_2 \\ \rho 
\end{pmatrix}.
\label{eq:eigenstates}
\end{equation}
Here $H_2$ is the SM-like Higgs, $h$, and $m_{H_{2}} \simeq m_h$ 
where the mixing angle $\theta$ between $H_2$ and $H_3$ is constrained as $\sin\theta\leq0.2$ by the LHC Higgs data \cite{Chpoi:2013wga, Cheung:2015dta,Cheung:2015cug} using the numerical analyses on the Higgs decay followed by \cite{Djouadi:1997yw, Djouadi:2006bz}.

{\tt CP-odd neutral scalar:} Calculating the mass matrix of the pseudo-scalars in a basis $(a_1, a_2, a_S)$ we get the mass matrix as  
\begin{equation}
M^2_A = \frac{\mu}{\sqrt{2}} \begin{pmatrix} \frac{v_2 v_S}{v_1} & - v_S & - v_2 \\ -v_S & \frac{v_1 v_S}{v_2} & v_1 \\
  -v_2 & v_1 & \frac{v_1 v_2}{v_S} \end{pmatrix} 
 \simeq  \begin{pmatrix} \frac{\mu v_2 v_S}{\sqrt{2} v_1} & 0 & 0 \\ 0 & 0 & 0 \\ 0 & 0 & 0 \end{pmatrix},
\end{equation}
using $S\simeq \frac{v_S+\rho+ i a_S}{\sqrt{2}}$. In the last step we used the approximation, $v_1 (\sim \mu) \ll v_2, v_S$.
 We find three mass eigenstates,
\begin{align}
A &= a_1 -\frac{v_1}{v_2} a_2 - \frac{v_1}{v_S} a_S, \nl
G^0 &= \frac{v_1}{v_2} a_1 +a_2,  \nl
a &= \frac{v_1}{v_S} a_1 -\frac{v_1^2}{v_2 v_S} a_2 +\left(1+\frac{v_1^2}{v_2^2} \right) a_S, 
\end{align}
up to normalization. They correspond to massive pseudo-scalar, the masslesss Nambu-Goldstone (NG) mode which is absorbed
by the $Z$ boson, and a massless physical Goldstone boson associated with the $U(1)_X$ breaking, respectively.
Hence the mass of $A$ is given by 
\begin{equation}
m_A^2 =\frac{\mu(v_1^2 v_2^2 + v_1^2 v_S^2 + v_2^2 v_S^2)}{\sqrt{2} v_1 v_2 v_S}  \simeq 
\frac{\mu v_2 v_S}{\sqrt{2} v_1},
\end{equation}
which is at the electroweak scale. It can be shown~\cite{Baek:2016wml}
that the Goldstone boson, $a$, is safe from the phenomenological
constraints such as $Z \to H_i a (i=1,2,3) $ decay, stellar cooling from the interaction $a  \overline{e} \gamma_5 e$,
{\it etc.}, because it interacts with the SM particles only via highly-suppressed ($\sim v_1/v_{2,S}$) mixing with the
SM Higgs.
 Note that, in our analysis below, we approximate pseudo-scalars as $A \simeq a_1$, $G^0 \simeq a_2$ and $a \simeq a_S$ since we assume $v_1 \ll v_2, v_S$ in realizing small neutrino mass.
Here we also discuss decoupling of the physical Goldstone boson from thermal bath where we assume it is thermalized via Higgs portal interaction. 
The interactions $ \rho \partial_\mu a_S \partial^\mu a_S/v_S$ , $\lambda_{2S} v_S v_2 \rho h_2$ and the SM Yukawa interactions 
generate the effective interaction among the Goldstone boson $a$ and the SM fermions
 \begin{equation}
 - \frac{\lambda_{2S} m_f}{2 m_{H_3}^2 m_{H_2}^2} \partial_\mu a \partial^\mu a \bar f f,
 \end{equation}
 where $m_f$ is the mass of the SM fermion $f$, and we used $a_s \simeq a$.
 The temperature, $T_a$, at which $a$ decouples from thermal bath is roughly estimated by~\cite{Weinberg:2013kea} 
 \begin{equation}
 \frac{\text{collision rate}}{\text{expansion rate}} \simeq \frac{\lambda_{2S}^2 m_f^2  T_a^5 m_{PL}}{m_{H_2}^4 m_{H_3}^4} \sim 1,
\label{eq:decoup_a}
 \end{equation}
 where $m_{PL}$ denotes the Planck mass and $m_f$ should be smaller than $T_a$ so that $f$ is in thermal bath. The decoupling temperature is then calculated by
 \begin{equation}
 T_a \sim 2 \, {\rm GeV} \left( \frac{m_{H_3}}{100 \, {\rm GeV}} \right)^{\frac{4}{5}} \left( \frac{{\rm GeV}}{m_f} \right)^{\frac{2}{5}} \left( \frac{0.01}{\lambda_{2S}} \right)^{\frac{2}{5}}.
 \end{equation}
 Thus Goldstone boson $a$ can decouple from thermal bath sufficiently earlier than muon decoupling and does not contribute to the effective
 number of active neutrinos\footnote{If $m_{H_3} \approx 500$ MeV and $\lambda_{2S} \approx 0.005$, then $a$ can make sizable 
contribution: $\Delta N_{\rm eff}=4/7$~\cite{Weinberg:2013kea}.}~\cite{Brust:2013xpv}.
 Note that the Goldstone boson should be in thermal bath at temperature below that of freeze-out of DM when we consider the relic density of
 DM, $X$, is explained by the process, $ X \bar{X} \to a a $, in our analysis below.
 Taking minimum DM mass as $\sim 100$ GeV freeze-out temperature $T_f$ is larger than $\sim 100/x_f$ GeV $\sim 4$ GeV where $x_f = m_{\rm DM}/T_f
 \sim 25$. Therefore we can get $T_f > T_a$ even with small $\lambda_{2S} (=0.01)$ as long as $m_{H_3}$ is not much heavier than the electroweak scale.
% in that case $m_{H_3}$ is preferred to be $\lesssim 30$ GeV to take $\lambda_{2S}$ to be small so that the scalar mixing can be sufficiently small. 

As the phenomenology of the Higgs sector has been discussed in \cite{Davidson:2009ha,Machado:2015sha, Bertuzzo:2015ada, Baek:2016wml},
we concentrate on the DM phenomenology in the following analysis.
%%%%%%%%%%%%%%%%%
\section{DM phenomenology}
\label{DMP}
%%%%%%%%%%%%%%%%%
In this section, we discuss DM physics of our model such as relic density, direct and indirect detections which are compared with experimental constraints. 
Since the Higgs portal interaction is strongly constrained by DM direct
detection~\cite{Baek:2011aa,Baek:2014kna,Baek:2012se,Cline:2013gha}, we consider the case of small mixing
so that $h_1 \simeq H_1$, $h_2 \simeq H_2$ and $\rho \simeq H_3$; here $H_2$ is the SM-like Higgs in our DM analysis.
%In our model $X$ is DM candidate.

%%%%%%%%%%%%%%%%%%
%%%%%%%%%%%%%%%%%%
\begin{figure}[t] 
\begin{center}
\includegraphics[bb=0 0 581 250, scale=0.23]{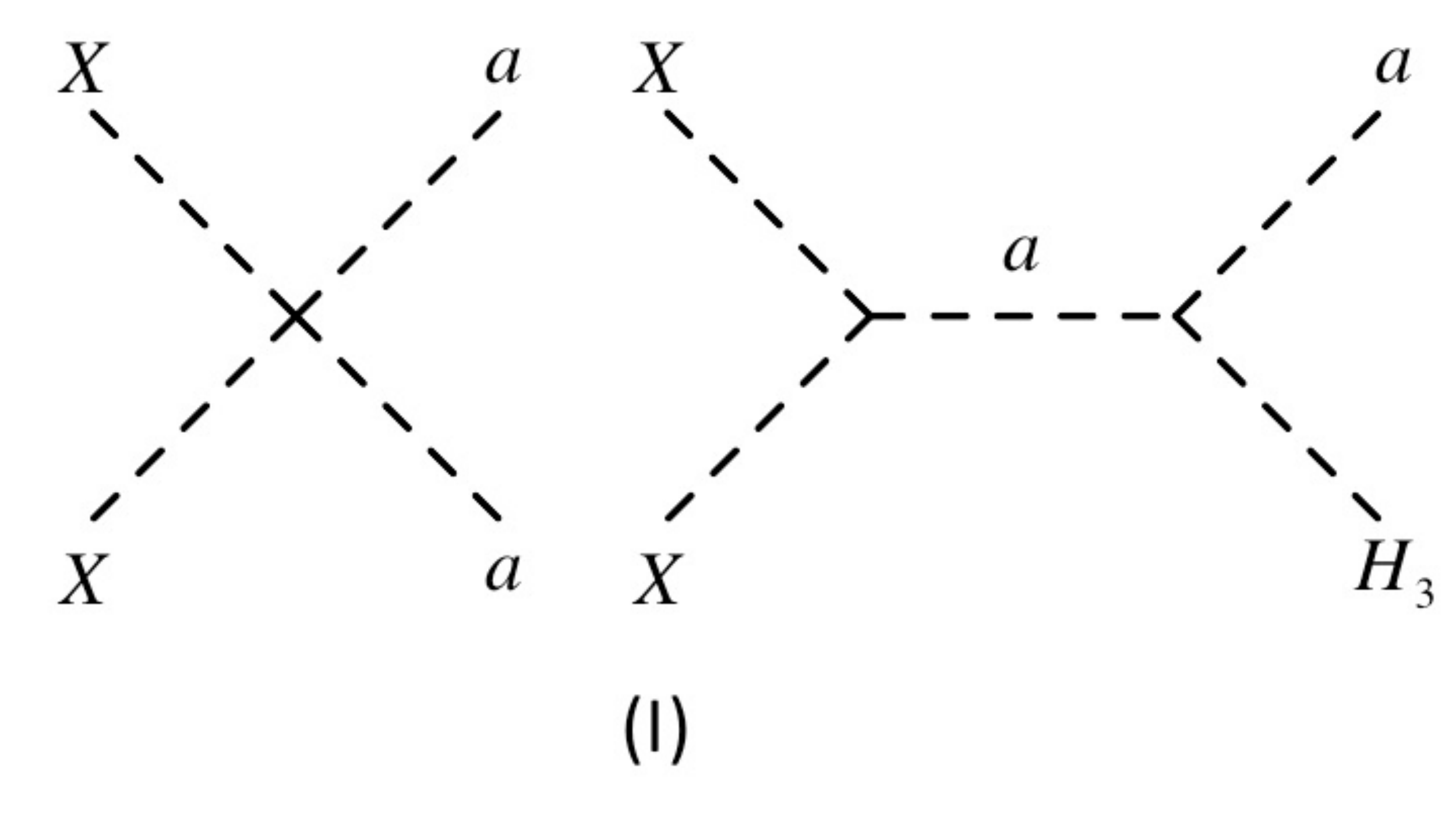} \qquad \qquad
\includegraphics[bb=0 0 581 250, scale=0.23]{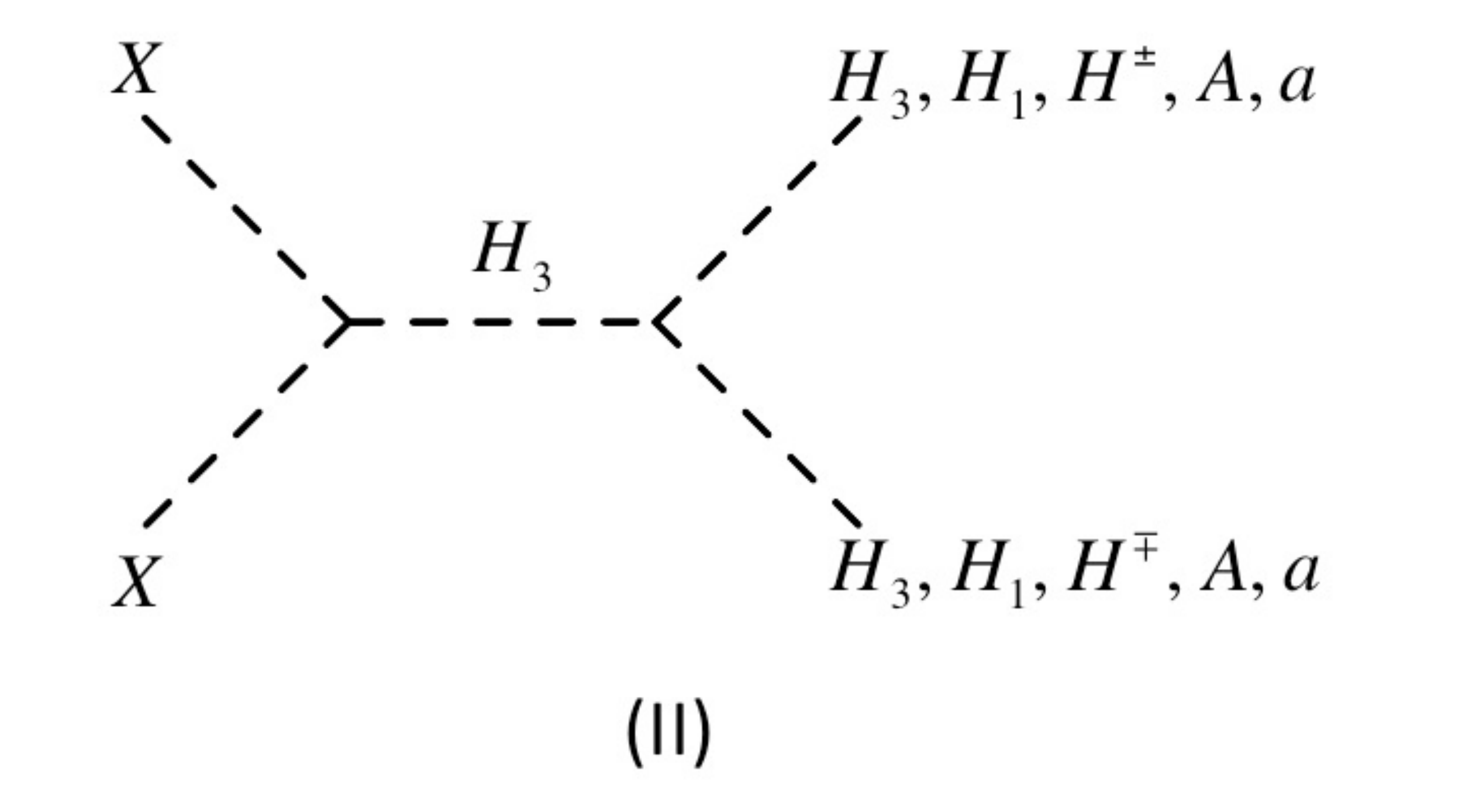}
\includegraphics[bb=0 0 650 450, scale=0.23]{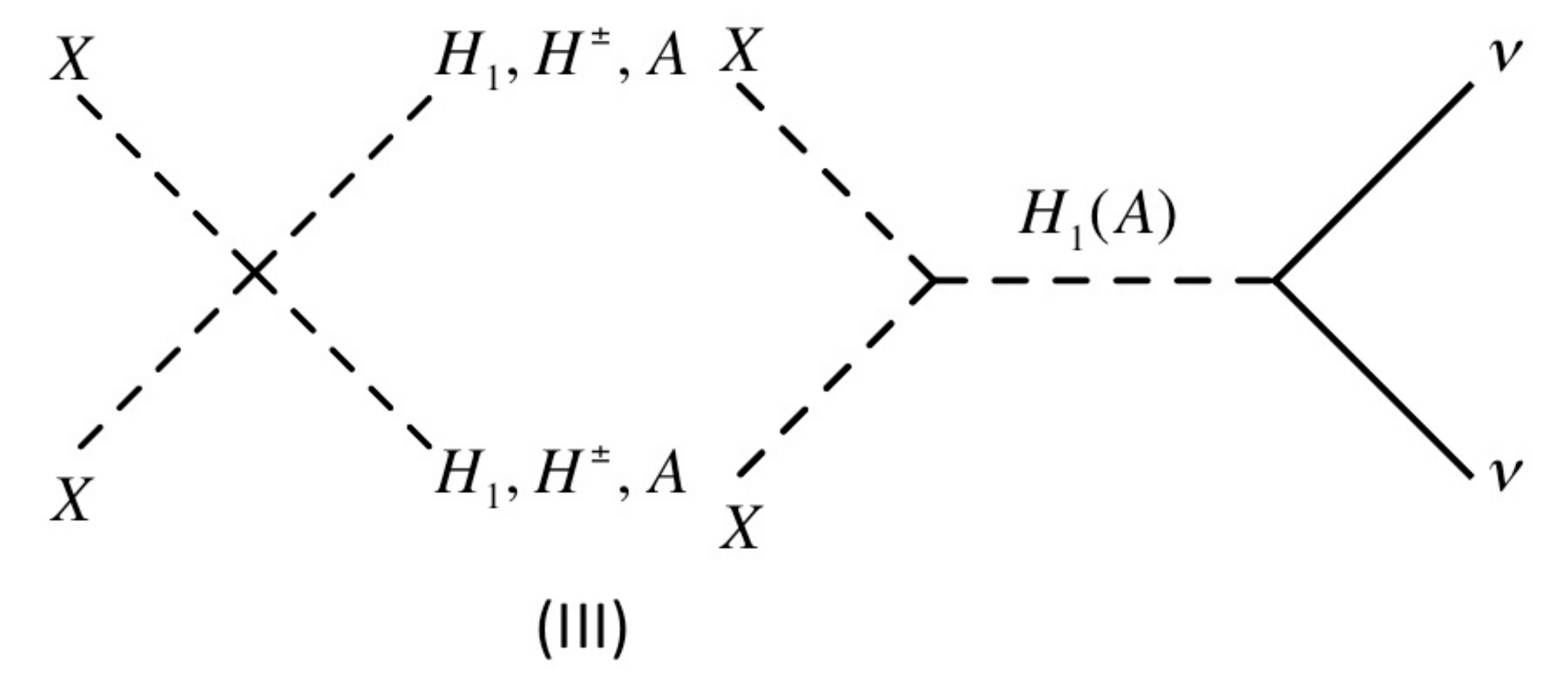} \qquad \qquad \qquad
\includegraphics[bb=0 0 450 450, scale=0.23]{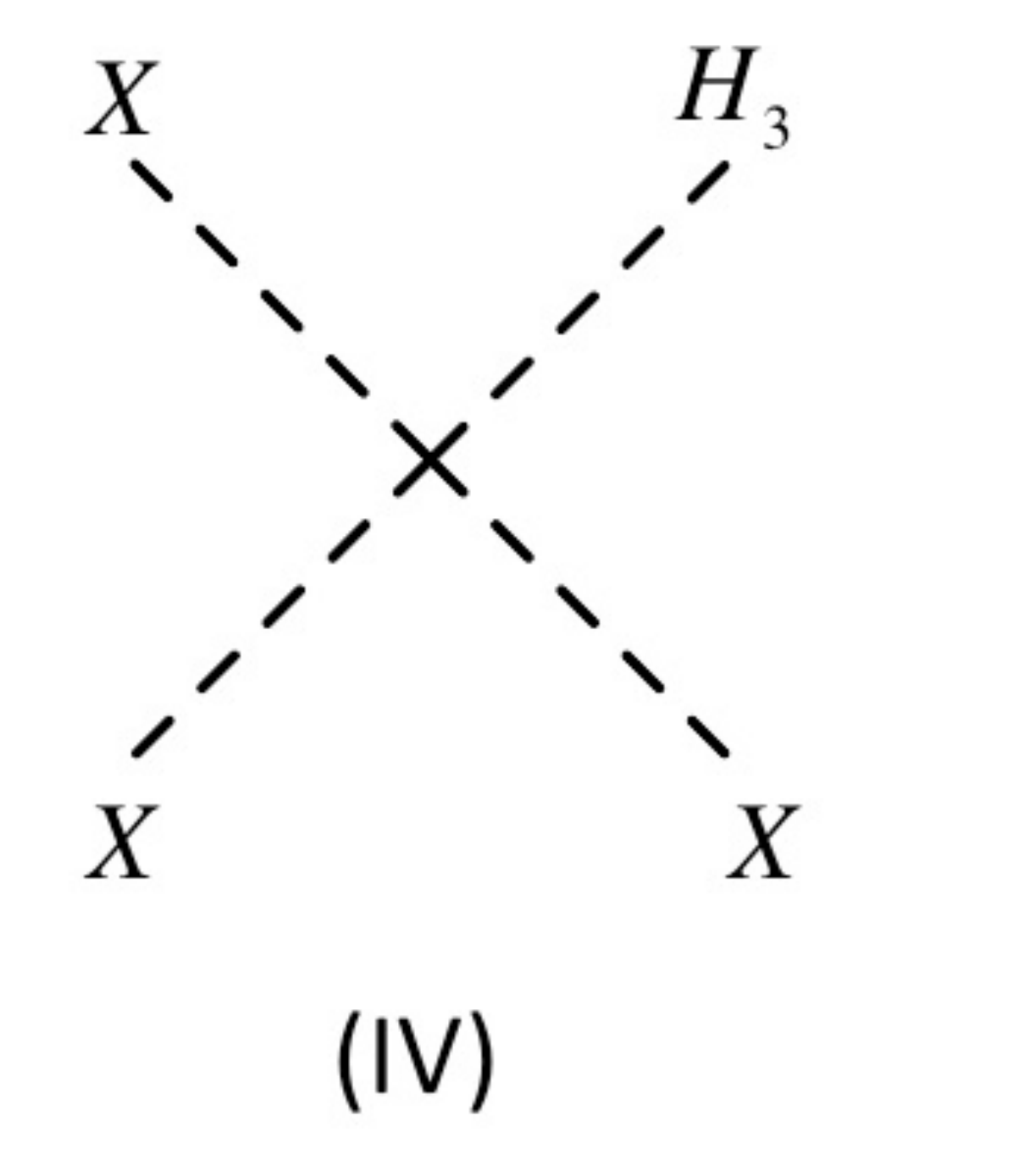}
\end{center}
\caption{Diagrams in (I), (II), (III) and (IV) correspond to DM annihilation process in scenario-I, II, III and IV. 
\label{fig:diagram1} }
\end{figure}
%%%%%%%%%%%%%%%%%%%%%%
\subsection*{Dark matter interaction}

Firstly masses of dark matter candidates $X$ is given by~\cite{Baek:2014kna} 
\begin{align}
m_{X}^2  = M_X^2 + \frac{\lambda_{1X}}{2} v_1^2 + \frac{\lambda_{2X}}{2} v_2^2 + \frac{\lambda_{SX}}{2} v_S^2  
%m_{X_I}^2 &= m_X^2 + \frac{\lambda_{1X}}{2} v_1^2 + \frac{\lambda_{2X}}{2} v_2^2 + \frac{\lambda_{SX}}{2} v_S^2 
\end{align} 
 where the real and imaginary part of $X$ has the same mass and $X$ is taken as a complex scalar field; this is due to remnant $Z_3$ symmetry. The interactions relevant to DM physics are given by 
\begin{align}
{\cal L} \ \supset \ & \frac{1}{v_S} \partial_\mu a (X \partial^\mu X^* - X^* \partial^\mu X) + \frac{1}{4 v_S^2} \partial_\mu a \partial^\mu a X^* X \nonumber \\
& + \frac{\lambda_{1X}}{2} \left(H^+ H^-  + \frac{1}{2} H_1^2 + A^2 \right)X^* X + \frac{\lambda_{2X}}{4} (2 v_2 H_2 + H_2^2)X^* X \nonumber \\
& + \frac{\lambda_{SX}}{4} (2v_S H_3 + H_3^2)X^* X +  \frac{\lambda_{3X}}{2} (v_S + H_3) (X X X + c.c.)  \nonumber \\
%& + \frac{\lambda_{12X}}{2} ( v_2 H_1 + H_1 H_2)(X_R^2 - X_I^2) - \lambda_{12X} (v_2 + H_2) A X_R X_I \nonumber \\ 
& - \mu_{SS} H_3^3 + \frac{1}{v_S} H_3 \partial_\mu  a \partial^\mu a  - \mu_{1S} H_3 \left(H^+ H^- + \frac{1}{2} (H_1^2 + A^2) \right) - \frac{\mu_{2S}}{2} H_3 H_2^2,
\label{eq:intDM}
\end{align} 

where we ignored terms proportional to $v_1$ since the value of VEV is tiny, $\mu_{SS} \equiv m_{H_3}^2/(2v_S)$,
$\mu_{1S} \equiv \lambda_{1S} v_S$, $\mu_{2S} \equiv \lambda_{2S} v_S$, and omitted scalar mixing $\sin \theta(\cos
\theta)$ assuming $\cos \theta \simeq 1$ and $\sin \theta \ll 1$.
Thus relevant free parameters to describe DM physics are summarized as;
\begin{equation}
\{ m_{X}, m_{H_1}, m_{H_3}, m_A, m_{H^\pm}, v_S, \lambda_{1X}, \lambda_{2X}, \lambda_{SX},  \lambda_{3X}, \mu_{1S}, \mu_{2S} \},
\end{equation}
where we choose $\mu_{1S,2S}$ as free parameter instead of $\lambda_{1S,2S}$ and we use $\mu_{SS} = m_{H_3}^2/(2v_S)$. 
In our analysis, we focus on several specific scenarios for DM physics by making assumptions for model parameters to
illustrate some particular processes of DM annihilations.
These scenarios are given as follows:
\begin{itemize}

\item Scenario-I: 100 GeV $< v_S < 2000$ GeV, $\{ \lambda_{1X}, \lambda_{2X}, \lambda_{SX}, \lambda_{3X}, \mu_{1S}/v \} \ll 1$.

\item Scenario-II :  $v_S \gg v$, $\{ \lambda_{SX}, \mu_{1S}/v \} \gg \{ \lambda_{1X}, \lambda_{2X}, \lambda_{3X}, \mu_{1S}/v \} $.

\item Scenario-III:  $v_S \gg v$,  $\lambda_{1X} \gg \{ \lambda_{2X}, \lambda_{SX}, \lambda_{3X}, \mu_{1S}/v \} $.

\item  Scenario-IV: $v_S \gg v$,  $\lambda_{X3} \gg \{\lambda_{1X}, \lambda_{2X}, \lambda_{SX}, \mu_{1S}/v \} $.

\end{itemize}
 Here we set $v\equiv v_2 \simeq 246\, {\rm GeV}$ since $v_1 \ll v_2$.
In scenario-I DM mainly annihilates into $a_S a_S$ and $a_S H_3$ final state as shown in Fig.~\ref{fig:diagram1}-(I). In scenario-II DM
annihilates via $H_3$ portal interaction as Fig.~\ref{fig:diagram1}-(II).
%; we take $\mu_{X}$ to be small and focus on coupling $\lambda_{SX}$  for simplicity.
In scenario-III DM annihilates into components of $\Phi_1$ through contact interaction with coupling
$\lambda_{1X}$ as shown Fig.~\ref{fig:diagram1}-(III). 
 Finally scenario-IV represents semi-annihilation processes  
  $X X \to X  H_3$ as shown in Fig.~\ref{fig:diagram1}-(IV).
In our analysis, we assumed $\lambda_{2S} \ll {\cal O}(1)$ so that we can neglect the case of DM annihilation via the SM
Higgs portal interaction since it is well known and constraints from direct detection experiments are strong.

%%%%%%%%%%%%%%%%%%
\begin{figure}[t] 
\begin{center}
\includegraphics[bb=-100 0 581 230, scale=1]{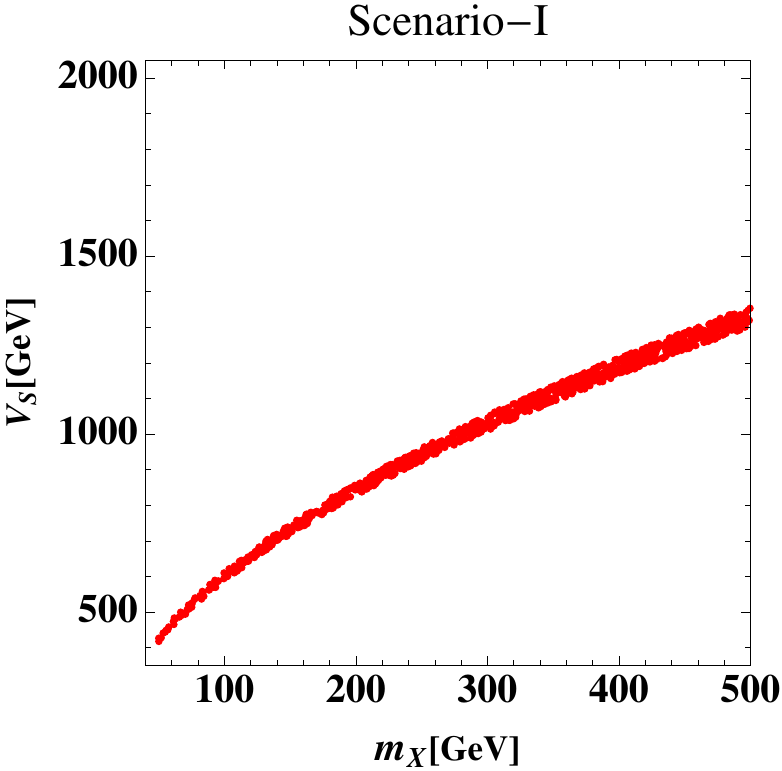} 
\end{center}
 \caption{ Scatter plot for parameters on $m_{X}$-$v_S$ plane under the DM relic abundance bound in Scenario-I.} 
\label{fig:DM1}
\end{figure}
%%%%%%%%%%%%%%%%%%%%%%
%%%%%%%%%%%%%%%%%%
\begin{figure}[t] 
\begin{center}
\includegraphics[bb=0 20 581 250, scale=0.94]{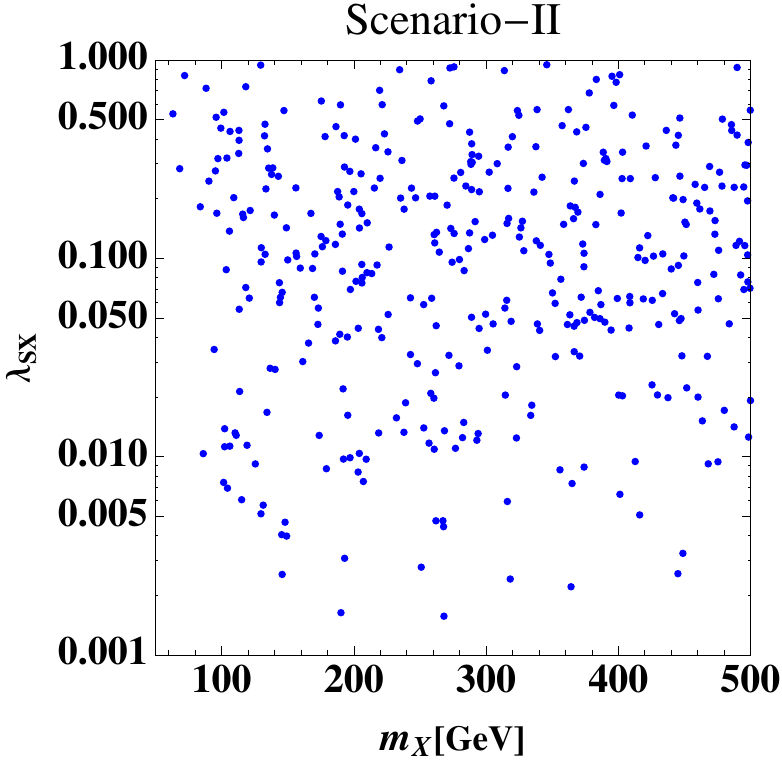} 
\includegraphics[bb=-300 0 581 0, scale=0.81]{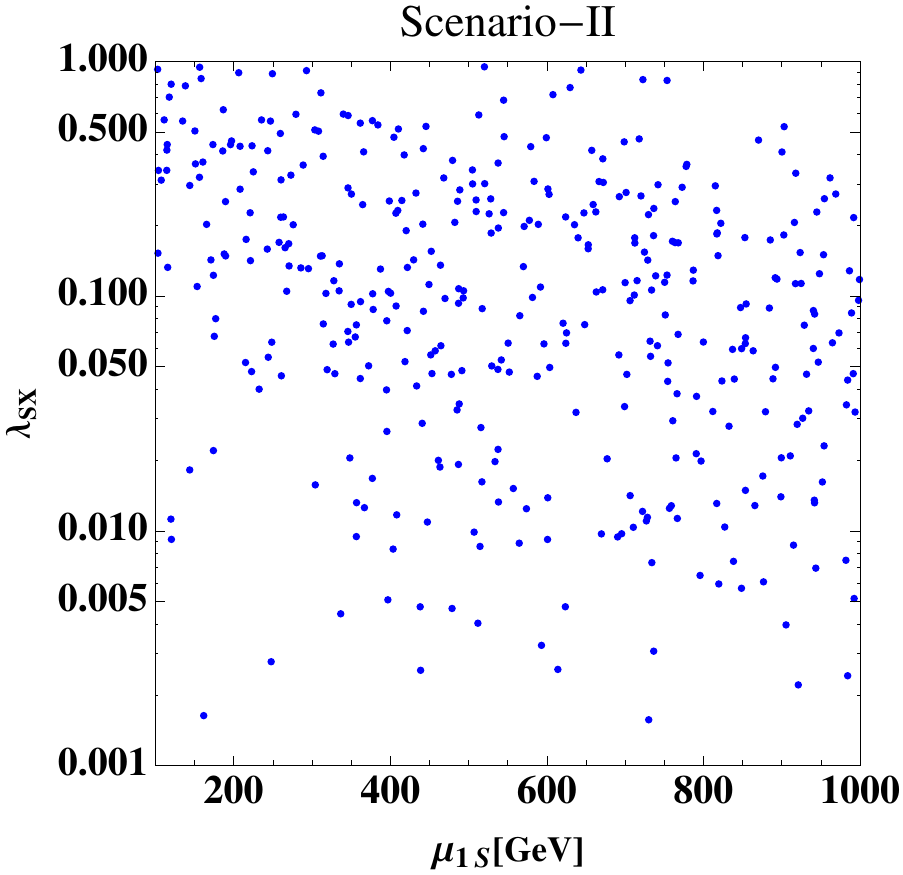} 
\end{center}
 \caption{ Scatter plot for parameters on $m_{X}$-$\lambda_{SX}$ and $\mu_{1S}$-$\lambda_{SX}$ planes in left and right panels under the DM relic abundance bound in Scenario-II.} 
\label{fig:DM2}
\end{figure}
%%%%%%%%%%%%%%%%%%%%%%
%%%%%%%%%%%%%%%%%%
\begin{figure}[t] 
\begin{center}
\includegraphics[bb=0 20 581 250, scale=0.94]{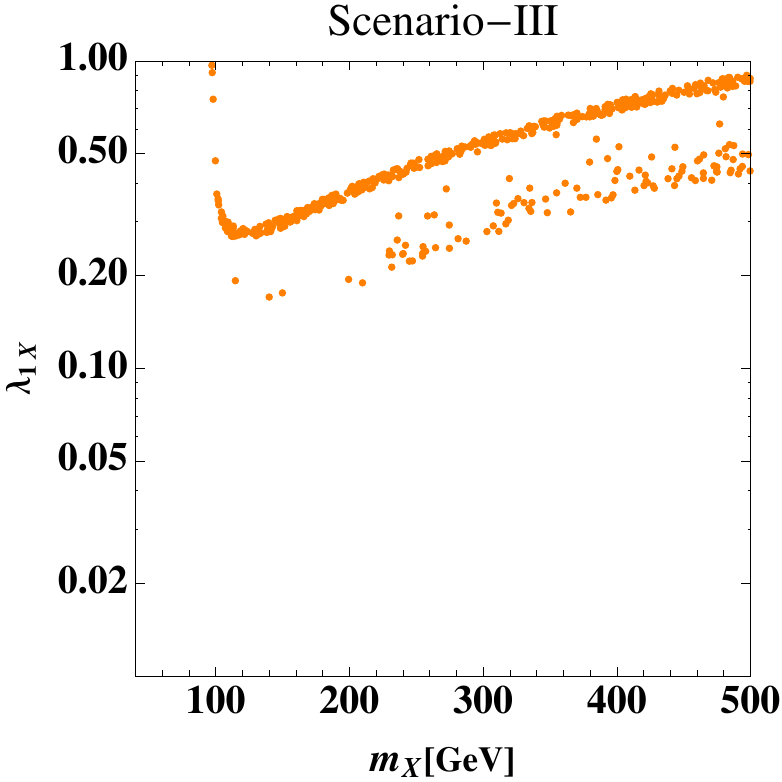} 
\includegraphics[bb=-230 0 581 0, scale=0.94]{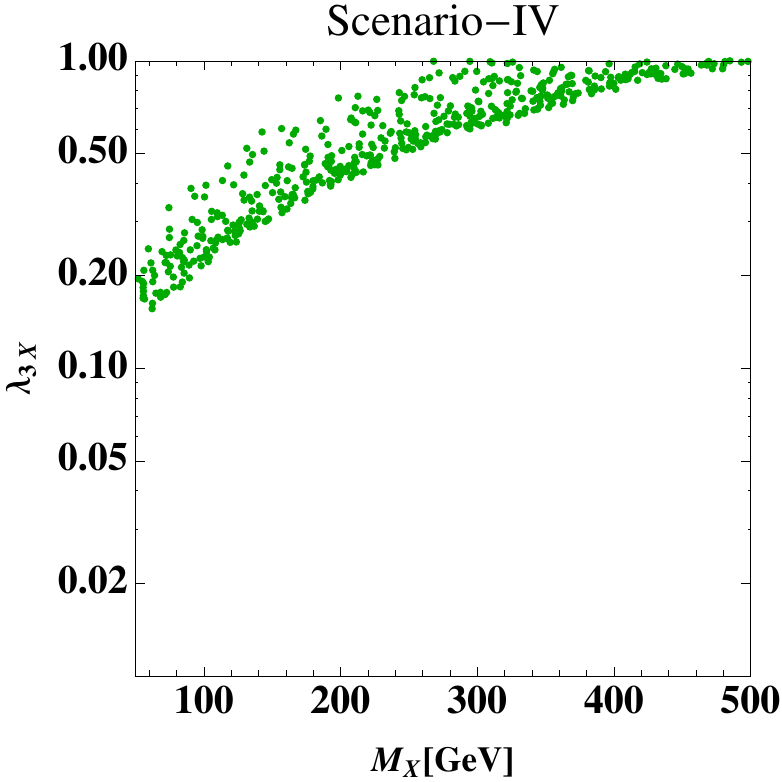}
\end{center}
 \caption{ Left: Scatter plot for parameters on $m_{X}$-$\lambda_{1X}$ plane under the DM relic abundance bound in Scenario-III. Right: that for parameters on $m_{X}$-$\lambda_{12X}$ in Scenario-IV} 
\label{fig:DM3}
\end{figure}
%%%%%%%%%%%%%%%%%%%%%%
\subsection*{Relic density}

Here we estimate the thermal relic density of DM for each scenario given above.
The relic density is calculated numerically with {\tt micrOMEGAs 4.3.5}~\cite{Belanger:2014vza} to solve the Boltzmann equation by implementing relevant interactions.
In numerical calculations we apply randomly produced parameter sets in the following parameter ranges.
For all scenarios we apply parameter settings as 
\begin{align}
&m_{X} \in [50,500] \ {\rm GeV}, \quad \mu_{2S} = 1 \ {\rm GeV}, \quad M_{H_1} = M_{A} = M_{H^\pm} \in [100, 1000] \ {\rm GeV},  \nonumber \\
& \lambda_{2X} \ll 1,
\end{align}
where the setting for $\lambda_{2X}$ is to suppress the SM Higgs portal interactions and small value of $\mu_{2S}$ is to suppress scalar mixing.
Then we set parameter region for each scenarios as follows:
\begin{align}
 {\rm Scenraio-I}: \ \ & \ v_S \in [100, 2000] \ {\rm GeV}, \quad \lambda_{SX, 1X, 3X} \in [10^{-8}, 10^{-4}], \nonumber \\
& \ \mu_{1S} \in [0.001, 0.1] \ {\rm GeV}, \quad  M_{H_3} \in [10, 30] \ {\rm GeV},  \\
 {\rm Scenario-II}: \ & \ v_S \in [3000, 10000] \ {\rm GeV}, \quad \lambda_{SX} \in [10^{-3}, 1], \quad \lambda_{1X, 3X} \in [10^{-8}, 10^{-4}], \nonumber \\
 & \ \mu_{1S} \in [100, 1000] \ {\rm GeV}, \quad  M_{H_3} \in [150, 2000] \ {\rm GeV}, \\
 {\rm Scenario-III}: & \ v_S \in [3000, 10000] \ {\rm GeV}, \quad \lambda_{1X} \in [10^{-3}, 1], \quad \lambda_{SX,  3X} \in [10^{-8}, 10^{-4}], \nonumber \\
 & \ \mu_{1S} \in [0.001, 0.1] \ {\rm GeV}, \quad  M_{H_3} \in [150, 2000] \ {\rm GeV}, \\ 
  {\rm Scenario-IV}: & \ v_S \in [3000, 10000] \ {\rm GeV}, \quad  \lambda_{3X} \in [10^{-3}, 1],  \quad \lambda_{SX, 1X} \in [10^{-8}, 10^{-4}],  \nonumber \\
 & \ \mu_{1S} \in [0.001, 0.1] \ {\rm GeV}, \quad  M_{H_3} \in [50, m_{X}] \ {\rm GeV}.
\end{align}
Then we search for the parameter sets which can accommodate with observed relic density.
Here we apply an approximated region~\cite{Ade:2015xua}
\begin{equation}
0.11 \lesssim \Omega h^2 \lesssim 0.13.
\end{equation}

In Fig.~\ref{fig:DM1}, we show parameter points on $m_{X}$-$v_S$ plane which can explain the observed relic density of DM in Scenario-I.
In this scenario, relic density is mostly determined by the cross section of $X X \to a_S a_S$ process which depends on $m_{X}/v_S$ via second term of the Lagrangian in Eq.~(\ref{eq:intDM}). Thus preferred value of $v_S$ becomes larger when DM mass increases as seen in Fig.~\ref{fig:DM1}.
In left and right panel of Fig.~\ref{fig:DM2}, we respectively show parameter points on $m_{X}$-$\lambda_{SX}$ and $\mu_{1S}$-$\lambda_{SX}$ planes satisfying correct relic density in Scenario-II.
In this scenario, the region $m_{X} \lesssim 100$ GeV requires relatively larger $\lambda_{SX}$ coupling since scalar boson modes $\{ H_3 H_3, H_1H_1, AA, H^\pm H^\mp \}$ are forbidden by our assumption for scalar boson masses. On the other hand the region $m_{X} > 100$ GeV allow wider range of $\lambda_{SX}$ around $0.01 \lesssim \lambda_{SX} \lesssim 1.0$  since DM can annihilate into other scalar bosons if kinematically allowed.
In left (right) panel of Fig.~\ref{fig:DM3}, we show parameter region on $m_{X}$-$\lambda_{1X}(\lambda_{ 3X} )$ satisfying the relic density in Scenario-III(IV).
In scenario-III,  DM mass should be larger than $\sim 100$ GeV to annihilate into scalar bosons from $\Phi_1$ and required value of the coupling is $0.2 \lesssim \lambda_{1X} \lesssim 1.0$ for $m_{X} \leq 500$ GeV. 
In  scenario-IV, the required value of the coupling $\lambda_{3X}$ has similar behavior as $\lambda_{1X}$ in the scenario-III for $m_X > 100$ GeV but slightly larger value.
This is due to the fact that semi-annihilation process require larger cross section than that of annihilation process.

\subsection*{Direct detection}

\begin{figure}[t] 
\begin{center}
\includegraphics[bb=-100 0 581 230, scale=0.95]{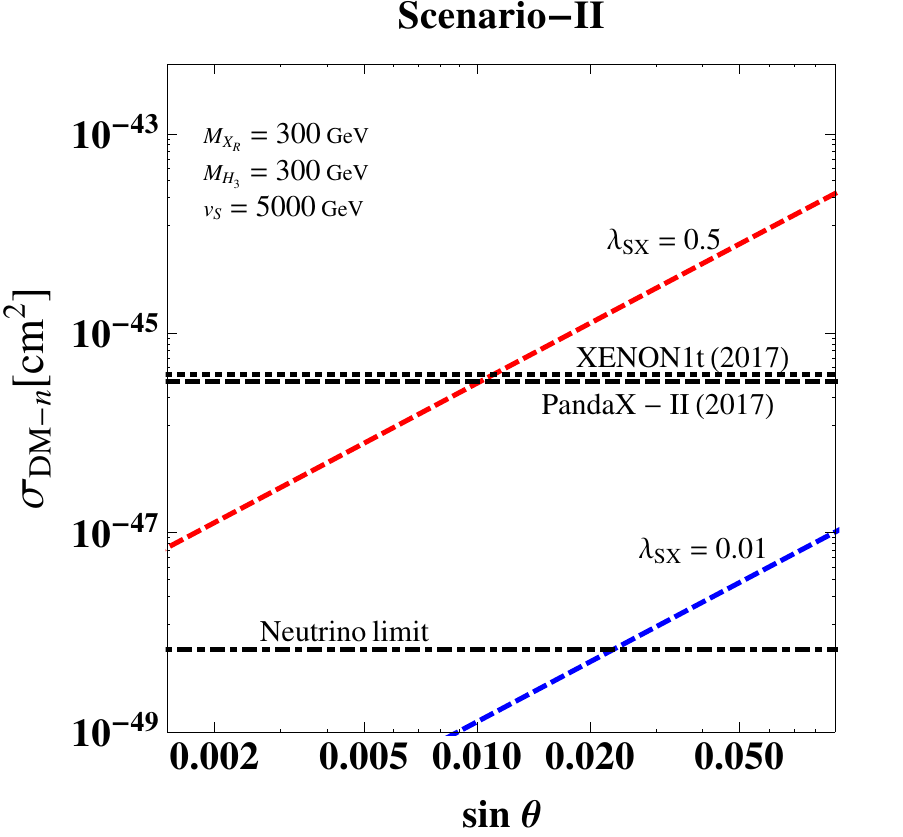}
\end{center}
 \caption{ DM-Nucleon scattering cross section as a function of $\sin \theta$ in Scenario-II where we take $m_{X} =
   300$ GeV, $m_{H_3}= 300$ GeV, $v_S = 5000$ GeV and $\lambda_{SX}=0.5(0.01)$ for red(blue) line as reference
   values. The current bounds from XENON1T~\cite{Aprile:2017iyp} and PandaX-II~\cite{Cui:2017nnn}}.  
\label{fig:DD}
\end{figure}

Here we briefly discuss constraints from direct detection experiments estimating DM-nucleon scattering cross section in our model.
Then we focus on our scenario-III since DM can have sizable interaction with nucleon via $H_2$ and $H_3$ exchange and investigate upper limit of mixing $\sin \theta$.
The relevant interaction Lagrangian with mixing effect is given by
\begin{equation}
\mathcal{L} \supset  \frac{\lambda_{SX} v_S}{2}  X^* X (c_\theta H_3 - s_\theta H_2) + \sum_{q} \frac{m_q}{v} \bar q q (s_\theta H_3 + c_\theta H_2),
\end{equation}
where $q$ denote the SM quarks with mass $m_q$, and we assumed $\mu_X \ll \lambda_{SX} v_S$ as in the relic density calculation.
We thus obtain the following effective Lagrangian for DM-quark interaction by integrating out $H_2$ and $H_3$;
\begin{equation}
\mathcal{L}_{\rm eff} = \sum_q \frac{\lambda_{SX} v_S m_q s_\theta c_\theta}{2v} \left( \frac{1}{m_h^2} - \frac{1}{m_{H_3}^2} \right) X^*X \bar q q,
\end{equation}
where $m_{H_2} \simeq m_h = 125$ GeV is used. The effective interaction can be rewritten in terms of nucleon $N$ instead of quarks such that
\begin{equation}
\mathcal{L}_{\rm eff} =  \frac{f_N  \lambda_{SX} v_S m_N s_\theta c_\theta}{v} \left( \frac{1}{m_h^2} - \frac{1}{m_{H_3}^2} \right) X^*X \bar N N,
\end{equation}
where $m_N$ is nucleon mass and $f_N$ is the effective coupling constant given by
\begin{equation}
f_N = \sum_q f_q^N = \sum_q \frac{m_q}{m_N} \langle N | \bar q q | N \rangle.
\end{equation}
The heavy quark contribution is replaced by the gluon contributions such that
\begin{align}
\sum_{q=c,b,t}  f_q^N = {1 \over m_N} \sum_{q=c,b,t} \langle N | \left(-{ \alpha_s\over 12 \pi} G^a_{\mu\nu}
  G^{a\mu\nu}\right) N \rangle,
\label{eq:f_Q}
\end{align}
which is obtained by calculating the triangle diagram for heavy quarks inside a loop.
Then we write the trace of the stress energy tensor as follows by considering the scale anomaly;
\begin{align}
\theta^\mu_\mu =m_N \bar{N} N = \sum_q m_q \bar{q} q - {7 \alpha_s \over 8 \pi} G^a_{\mu\nu} G^{a\mu\nu}.
\label{eq:stressE}
\end{align}
Combining Eqs.~(\ref{eq:f_Q}) and (\ref{eq:stressE}), we get 
\begin{align}
\sum_{q=c,b,t} f_q^N = \frac{2}{9} \left( 1 - \sum_{q = u,d,s} f_q^N \right),
\end{align}
which leads
\begin{align}
f_N = \frac29+\frac{7}{9}\sum_{q=u,d,s}f_{q}^N.
\end{align}
Finally we obtain the spin independent $X$-$N$ scattering cross section as follows;
\begin{equation}
\sigma_{\rm SI}(X N \to X N) =  \frac{1}{8 \pi} \frac{\mu_{NX}^2 f_N^2 m_N^2 \lambda_{SX}^2 v_S^2 s_\theta^2 c_\theta^2}{v^2 m_{X}^2} 
 \left( \frac{1}{m_h^2} - \frac{1}{m_{H_3}^2} \right)^2,
\end{equation}
where $\mu_{NX} = m_N m_{X}/(m_N + m_{X})$ is the reduced mass of nucleon and DM.
Here we consider DM-neutron scattering cross section for simplicity where that of DM-proton case gives almost similar result.
In this case, we adopt the effective coupling $f_n \simeq 0.287$ (with $f_u^n = 0.0110$, $f_d^n = 0.0273$, $f_s^b = 0.0447$) in estimating the cross section.
In Fig.~\ref{fig:DD}, we show DM-nucleon scattering cross section as a function of $\sin \theta$ we take $m_{X} = 300$ GeV, $m_{H_3}= 300$ GeV, $v_S = 5000$ GeV, and $\lambda_{SX}= 0.5(0.01)$ for red(blue) line as reference values. We find that some parameter region is constrained by direct detection when $\lambda_{SX}$ is relatively large and $\sin \theta > 0.01$.
More parameter region will be tested in future direct detection experiments.

%%%%%%%%%%%%%%%%%%
%%%%%%%%%%%%%%%%%%%%%%

The Higgs portal interaction can be also tested by collider experiments.
The interaction can be tested via searches for invisible decay of the SM Higgs for $2 m_X < m_h$ while 
collider constraint is less significant compared with direct detection constraints for $2m_X > m_h$~\cite{Khachatryan:2016whc, Hoferichter:2017olk,Aad:2015pla}.
Furthermore DM can be produced via heavier Higgs boson $H_3$ if $2 m_X < m_{H_3}$ and the possible signature will be mono-jet with missing transverse momentum as $pp \to H_3 j \to XX j$.
However the production cross section will be small when the mixing effect $\sin \theta$ is small as we assumed in our analysis.
Such a process would be tested in future LHC with sufficiently large integrated luminosity while detailed analysis is beyond the scope of this paper.

\subsection*{Indirect detection}

Here we discuss possibility of indirect detection in our model by estimating thermally averaged cross section in current Universe with {\tt micrOMEGAs 4.3.5} using allowed parameter sets from relic density calculations. Since $a_Sa_S$ final state is dominant in scenario-I, we focus on the other scenarios in the following. 

\begin{figure}[t] 
\begin{center}
\includegraphics[bb=0 20 581 280, scale=0.8]{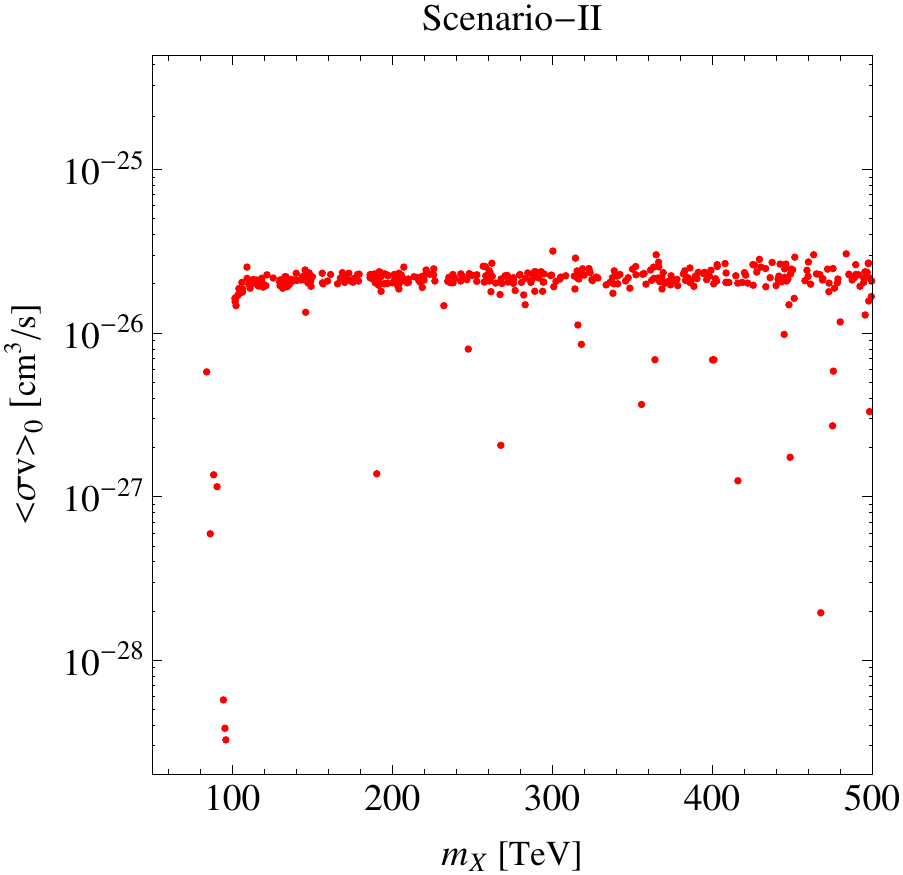}
\includegraphics[bb=-280 0 581 20, scale=0.8]{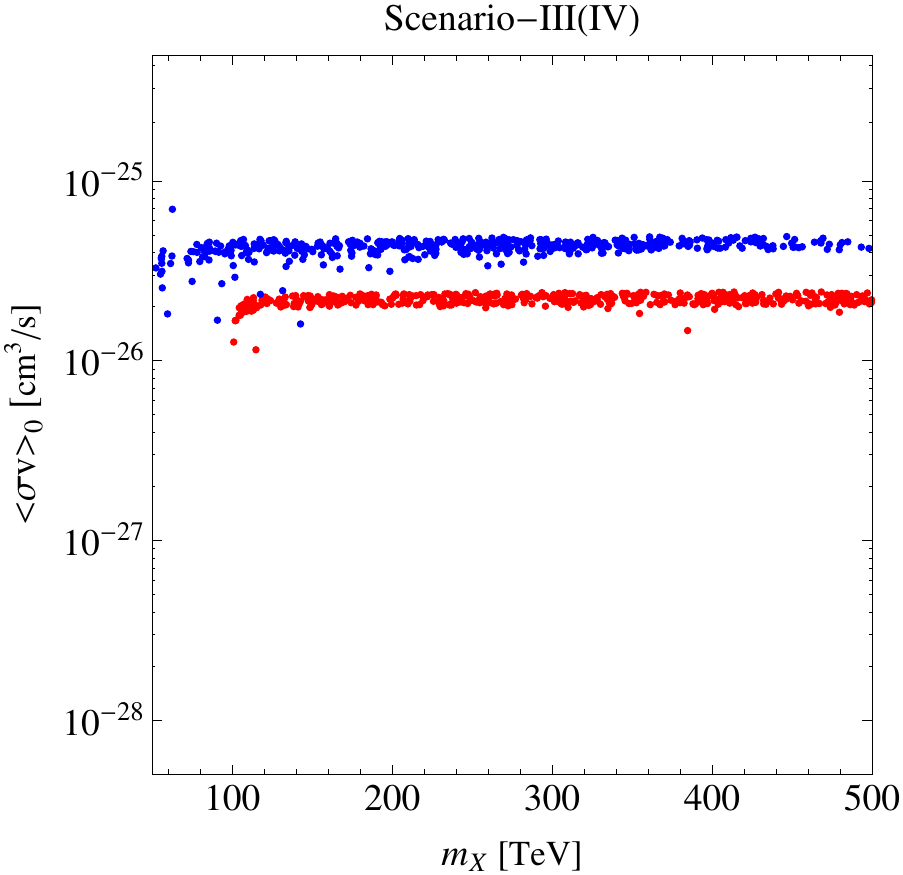}
\end{center}
 \caption{ Left: the current DM annihilation cross section in Scenario-II as a function of $m_{X}$. Right: that for Scenario-III and IV represented by red and blue points.} 
\label{fig:ID}
\end{figure}

Fig.~\ref{fig:ID} shows DM annihilation cross section in current Universe as a function of $m_{X}$ where left and right panels correspond to Scenario-II and Scenario-III/IV.
In Scenario-II, the cross section is mostly $\sim O(10^{-26})$cm$^{-3}/$s while some points give smaller(larger) values corresponding to the region with $2 m_{X} \gtrsim(\lesssim) M_{H_3}$ as a consequence of resonant effect.
The annihilation processes in the scenario provide the SM final state via decay of $H_3$ and $\{H_1, H^\pm, A\}$ where $H_3$ decay gives mainly $b \bar b$ via mixing with the SM Higgs and the scalar bosons from second doublet gives leptons. This cross section would be tested via $\gamma$-ray observation like Fermi-LAT~\cite{Ackermann:2013yva} as well as high energy neutrino search such as IceCube~\cite{Aartsen:2015zva, Aartsen:2017mau}, especially when the cross section is enhanced.
In Scenario-III, the cross section is mostly $\sim O(10^{-26})$cm$^{-3}/$s and the final states from DM annihilation include components of $\Phi_1$ that are $\{H_1, H^\pm, A\}$. Thus DM mainly annihilate into neutrinos via the decay these scalar bosons while little amount of charged lepton appear from $H^\pm$. Therefore constraints from indirect detection is weaker in this scenario.
 In Scenario-IV, the values of cross section is relatively larger due to the nature of semi-annihilation scenario.
In this case final states from DM annihilation give mostly $b \bar b$ via decays of $H_3$ in the final state. Then it would be tested by $\gamma$-ray search and neutrino observation as in the scenario-II. 

%%%%%%%%%%%%%%%%%%
%%%%%%%%%%%%%%%%%%%%%%

%%%%%%%%%%%%%%%%%
\section{Conclusion}
\label{Conc}
%%%%%%%%%%%%%%%%%
We consider a neutrino Two Higgs Doublet Model ($\nu$THDM) in which small Dirac neutrino masses are explained 
by small VEV, $v_1 \sim {\cal O}(1)$ eV, of Higgs $H_1$ associated with neutrino Yukawa interaction. A global $U(1)_X$
symmetry is introduced to forbid seesaw mechanism. The smallness of $v_1$ proportional to soft $U(1)_X$-breaking
parameter $m_{12}^2$ is technically natural.

We extend the model to introduce a scalar dark matter candidate $X$ and scalar $S$ breaking $U(1)_X$ symmetry
down to discrete $Z_2$ symmetry. Both are charged under $U(1)_X$. The lighter state of $X$ is stable since it is the
lightest particle with $Z_2$ odd parity. The soft parameter $m_{12}^2$ is replaced by $ \mu \langle S \rangle$.
The physical Goldstone boson whose dominant component is pseudoscalar part of $S$ is shown to be phenomenologically
viable due to small ratio ($\sim {\cal O}(10^{-9})$) of $v_1$ compared to electroweak scale VEVs of the SM Higgs and
$S$.

We study four scenarios depending on dark matter annihilation channels in the early Universe to simplify the analysis of
dark matter phenomenology. In Scenario I, Goldstone modes are important. Scenario II is $H_3$ portal. In Scenario III,
the dark matter makes use of the portal interaction with $\Phi_1$ which generates Dirac neutrino masses. 
 In Scenario IV the dominant interaction is $\lambda_{3X} S^\dagger X X X + h.c.$ which induces semi-annihilation process of our dark matter candidate.
In Scenario II, the dark matter scattering cross section with neucleons can be sizable and detected at next generation
direct detection experiments. We calculated indirect detection cross section in Scenarios II, III, and IV, which can be
tested by observing cosmic $\gamma$-ray and/or neutrinos.

%%%%%%%%%%%%%%%%%%%%%%%%%%%%%%%%%%%
%\vspace{0.5cm}
%\hspace{0.2cm} {\bf Acknowledgments}
\section*{Acknowledgments}
\vspace{0.5cm}
This work is supported in part by National Research Foundation of Korea (NRF) Research Grant NRF-2015R1A2A1A05001869 (SB).
%%%%%%%%%%%%%%%%%%%%%%%%%%%%%%%%%%%

%%%%%%%%%%%%%%%%%%%%%%%%%%%%%%%%%%%%%%%%%%%%%%%%%%%%%%%%%%%%%%%%%%%%%%%%%%%%%%
\providecommand{\href}[2]{#2}
\addcontentsline{toc}{section}{References}
\bibliographystyle{JHEP}
\bibliography{BDN_rev}

\end{document}